\documentclass[11pt,letterpaper]{llncs}
\usepackage{soul}
\usepackage[utf8]{inputenc}
\usepackage[table]{xcolor}
 \usepackage{cancel} 
\usepackage{float}
\usepackage[margin=1in]{geometry}
\usepackage{graphicx}
\usepackage{xcolor}
\usepackage{amsmath}
\usepackage{amssymb}
\usepackage[sort,nocompress]{cite}
\usepackage[ruled,vlined,lined,boxed,commentsnumbered]{algorithm2e}
\usepackage{todonotes}
\usepackage{verbatim}
\usepackage{subcaption}
\usepackage[normalem]{ulem}
\usepackage{array}
\usepackage{algorithmic}
\usepackage{mathrsfs}
 \usepackage{multirow}
 \usepackage{blindtext}
\usepackage{hyperref}

\usepackage{todonotes}

\usepackage{pgf}
\usepackage{tikz}

\SetKwComment{Comment}{$\triangleright$\ }{}
\newcommand{\notaN}[1]{\textcolor{orange}{#1 NP}}

\newcommand{\patlen}{\pi}
\newcommand{\aut}{\mathcal A}

\usetikzlibrary{arrows,automata,positioning}
\title{Co-lexicographically Ordering Automata and Regular Languages - Part II}

\author{
Nicola Cotumaccio\inst{1, 2}, Giovanna D'Agostino\inst{3}, Alberto Policriti\inst{3}, Nicola Prezza\inst{4}
} 

\institute{
Gran Sasso Science Institute, L'Aquila,  Italy. Email: \email{nicola.cotumaccio@gssi.it} \and Dalhousie University, Halifax, Canada. Email: \email{nicola.cotumaccio@dal.ca} \and University of Udine, Italy.  Email: \email{giovanna.dagostino@uniud.it, alberto.policriti@uniud.it} \and University Ca' Foscari, Venice,  Italy. Email: \email{nicola.prezza@unive.it}
}

\date{\today}

\begin{document}

\maketitle
\thispagestyle{empty}

\begin{abstract}

Consider the problem of matching a pattern $P$ of length $|P|$ against the elements of a given regular language $\mathcal L$ in the setting where  $\mathcal L$ can be \emph{pre-processed off-line} in a fast data structure (an index). \emph{Regular expression matching} is an ubiquitous problem in computer science, finding fundamental applications in areas including, but not limited to, natural language processing, search engines, compilers, and databases. 
Recent results have settled the exact complexity of this problem:  $\Theta(|P| m)$ time is necessary and sufficient for indexed pattern matching queries, where $m$ is the size of an NFA recognizing $\mathcal L$.
This, however, does not mean that \emph{all} regular languages are hard to index: for instance, for the sub-class of \emph{Wheeler languages} \cite{alanko2020wheeler} we can reduce query time to the optimal $O(|P|)$. 
A Wheeler language admits a \emph{total} order of a finite refinement of its Myhill-Nerode equivalence classes reflecting the co-lex order of their elements. This boosts indexing performance because classes whose elements are suffixed by $P$ form a \emph{range} in this order.
In \cite{cotumaccio2021indexing} this technique was extended to \emph{arbitrary} NFAs by allowing the order to be \emph{partial}. 
This line of attack suggested that the \emph{width $p$} of such an order is the parameter ultimately capturing the fine-grained complexity of the problem:

(i) indexed pattern matching can always be solved in $\tilde O(|P| p^2)$ time,  (ii) the classic powerset construction algorithm always produces an output whose size is exponential in $p$ rather than in the input's size, and (iii) $p$ even determines how succinctly NFAs can be encoded.

In the present work, 
we tackle the regular language indexing problem by first studying the hierarchy of \emph{$p$-sortable languages}: regular languages accepted by automata of width $p$. We show that the hierarchy is strict and does not collapse, and provide (exponential in $p$) upper and lower bounds relating the minimum widths of equivalent NFAs and DFAs. 
Our bounds indicate the importance of being able to index NFAs, as they enable indexing regular languages with much faster and smaller indexes. Our second contribution solves precisely this problem, optimally: we devise a polynomial-time algorithm that indexes any NFA  with the optimal value $p$ for its width, \emph{without} explicitly computing $p$ (NP-hard to find). In particular, this implies that we can index in polynomial time the well-studied case $p=1$ (Wheeler NFAs).
More in general,  in polynomial time we can build an index breaking the worst-case conditional lower bound of $\Omega(|P| m)$, whenever the input NFA's width is $p \in o(\sqrt{m})$.

\end{abstract}

\newpage

\setcounter{page}{1}

\section{Introduction}


String indexing is the algorithmic problem of building a small data structure (an \emph{index}) over a given string supporting fast substring search queries \cite{NavarroIndexingSurvey}.
Building efficient string indexes is a challenging problem finding important applications in several areas, notably bioinformatics \cite{li2009fast,langmead2009ultrafast}. Lifting this problem to a \emph{regular collection} $\mathcal L$ of strings is an even more challenging problem and naturally calls into play finite state automata.
As a matter of fact, \emph{Regular expression matching} is an ubiquitous problem in computer science, finding fundamental applications in areas including, but not limited to, natural language processing, search engines, compilers, and databases.
When $\mathcal L$ is represented as an NFA (equivalently, a regular expression) of size $m$, existing on-line algorithms \cite{hypertext_amir} solve the problem in $O(\patlen m)$ time, $\patlen$ being the length of the query pattern.  Recent lower bounds by Backurs and Indyk \cite{backurs2016regular}, Equi et al.~\cite{equi2020conditional,equi19}, Potechin and Shallit \cite{potechin2020lengths}, and Gibney~\cite{gibney2020simple} show that, unless important conjectures such as the Strong Exponential Time Hypothesis (SETH) \cite{Impagliazzo01} fail, this complexity cannot be significantly improved. This holds even in the \emph{off-line} setting (the subject matter of our work) where $\mathcal L$ can be pre-processed in an index in polynomial time and the complexity is measured in terms of query times \cite{equi2020graphs}. As pointed out by Backurs and Indyk  \cite{backurs2016regular}, Gagie et al. \cite{GAGIE201767}, Alanko et al. \cite{alanko2020regular,alanko2020wheeler}, and Cotumaccio and Prezza \cite{cotumaccio2021indexing}, however, this does not necessarily mean that \emph{all} regular languages are hard to index.

Indeed, \cite{alanko2020regular,alanko2020wheeler}
tackled the task of
characterising regular languages admitting a direct generalization of known string indexing techniques---those accepted by so-called \emph{Wheeler automata} introduced in \cite{GAGIE201767}.
More specifically, recalling that a state $ q $ of an NFA  can be seen as the collection $ I_{q} $  of strings 
labeling the paths that connect the start state with $q$, 
\cite{alanko2020regular,alanko2020wheeler} showed that  Wheeler automata are those for which (i) each $I_q$  is a \emph{convex} set  in the co-lexicographic (for brevity, co-lex) ordered set of 
strings read on the automaton's paths, and (ii) the family of these $I_q$ enjoys the so-called \emph{prefix/suffix property}: the only way $I_q$ can  intersect another $I_{q'}$ is that a suffix of the former coincides with a prefix of the latter (or vice versa). 
 
In other words, the co-lex order over strings can be naturally lifted to the elements of a family of convex sets   enjoying such property. In turn, this defines an order over the automaton's states which enables pattern matching queries in \emph{optimal} $O(\pi)$ time: states reached by a path labeled with a given string $P$ form an \emph{interval} in this order \cite{GAGIE201767}.

\medskip

Since, clearly, not all (interesting, regular) languages admit a Wheeler accepting automaton, the next natural question is: what if we want to index a general regular language? can we say something on the language's \emph{propensity}  to be indexed? can we give directions/bounds on the complexity of such indexing task? 

In this paper, elaborating on the idea put forward in \cite{cotumaccio2021indexing}, we prove that the above ideas are a sort of one-dimensional version of a more general setting. 
From this more general point of view,  the set of the $I_q$'s is (always, for \emph{any} automaton) \emph{partially ordered} and all its elements end up in a collection of $ p $ totally ordered components, where $p$ is the order's width.

It turns out that the order's width $p$ is a fundamental measure of NFA complexity \cite{cotumaccio2021indexing}: (i) indexed pattern matching can always be solved in $\tilde O(\patlen p^2)$ time (the Wheeler case corresponding to $p=1$),  (ii) the standard powerset construction algorithm always produces an output whose size is exponential in $p$, rather than in the input's size, and (iii) $p$ even determines how succinctly NFAs can be encoded ($O(1+\log p)$ bits per edge, assuming constant alphabet for simplicity).
  
  Within this framework, our main contribution
  is to begin the study of the hierarchy (encompassing all regular languages) of \emph{$p$-sortable languages} --- regular languages accepted by automata of width $p$ (for the minimum such $p$) --- and show how to efficiently index them. In this hierarchy, regular languages are sorted according to the new measure of NFA complexity, the width.
  More in detail:
  
  \begin{enumerate}
      \item[(1)] We show that the hierarchy is strict and does not collapse: a language of width $ p $ exists for all $p\geq 1$.
      \item[(2)] We explore the effect that determinism has on the automaton's width. We prove upper- and lower-bounds showing that determinism forces an \emph{exponentially-large} $p$ in the worst case.
      \item[(3)] While in \cite{cotumaccio2021indexing} it was shown how to index DFAs in polynomial time for the optimal value of $p$, our contribution (2) suggests that this is not yet sufficient in order to index regular languages optimally. We therefore devise a polynomial-time algorithm that indexes any NFA for its optimal value $p$ of width. 
  \end{enumerate}
  
  Crucially, contribution (3) is achieved in a \emph{width-oblivious} sense, that is, without explicitly computing $p$. Notice that a detour is unavoidable, since computing the smallest $p$ of an NFA is an NP-hard problem \cite{cotumaccio2021indexing,gibney2019ESA}. 

Our findings have important algorithmic consequences for the regular expression matching problem: any NFA $\mathcal A$ of size $m$ and width $p\in o(\sqrt{m})$ can be indexed in polynomial time so that pattern matching queries on $\mathcal L(\mathcal A)$ (for example, membership) can be solved in $o(\pi m)$ time. 
This breaks  the conditional lower bound $\Omega(\pi  m)$ of Equi et al. \cite{equi2020graphs} holding in the worst-case even when polynomial preprocessing time is allowed.   
A particular case of interest is the well-studied case of $p=1$: Wheeler NFAs \cite{GAGIE201767}. While this class of NFAs supports pattern matching queries in optimal $O(\pi)$ time, deciding membership of an NFA to this class (and thus indexing all and only the Wheeler NFAs) was proven to be NP-complete already in \cite{gibney2019ESA}. Our index-construction algorithm sidesteps this problem by indexing a \emph{strictly larger} class of NFAs, thus not directly deciding Wheelerness of the input NFA. 
 
 The paper is organized as follows: after giving some definitions and basic results in Section \ref{sec:colex orders}, in Section \ref{sec:width} we discuss the notion of \emph{width} of a regular languages and relate the two hierarchies---deterministic/nondeterministic---based on this notion. Finally, in Section \ref{sec:indexing} we exhibit a polynomial-time algorithm indexing NFAs with the optimal value for their width $p$. In order to achieve this, in Subsection \ref{sec:sorting} we first exhibit a new order on the states of an NFA that is strictly more general than co-lex orders and that can be computed in polynomial time. In Subsection \ref{subsec:indexing} we show the states of an NFA, when sorted according to our order, enable solving fast pattern matching queries on the NFA's paths.

 Due to limited space, the proofs can be found in the appendix.
 
\section{Notation}\label{sec:notation}
We say that $ (V, \le) $ is a {\em partial order} if $ V $ is a set and $ \le $ is a binary relation on $ V $ being reflexive, antisymmetric and transitive. Any $  u, v \in V $ are said to be $ \le $-{\em comparable} if either $ u \le v $ or $ v \le u $ hold. 
We write $ u < v $ when $ u \le v $ and $ u \not = v $. We write $ u ~\|~ v $ if $ u $ and $ v $ are not  $\leq$-comparable. Note that for every $ u, v \in V $ exactly one of the following hold true: (1) $ u = v $, (2) $ u < v $, (3) $ v < u $, (4) $ u ~\|~ v $. We say that $ (V, \le) $ is a {\em total order} if $ (V, \le) $ is a partial order and every pair of elements in $(V, \le)$ are $\le$-comparable.   A subset $ Z \subseteq V $ is a $\le$-{\em chain} if $ (Z, \le) $ is a total order, and a family $ \{V_i\}_{i = 1}^p $ is a $\le$-{\em chain partition} if $ \{V_i\}_{i = 1}^p $ is a partition of $ V $ and each $ V_i $ is a $\le$-chain. The {\em width} of $ (V, \le) $ is the smallest integer $ p $ for which there exists a chain partition $ \{V_i \}_{i = 1}^p $. We say that $ U \subseteq V $ is an $\le$-{\em antichain} if every pair of elements in $ U $ are not $ \le $-comparable. Dilworth's theorem \cite{dilworth2009decomposition} states that the width of $ (V, \le) $ is the cardinality of a largest $\le$-antichain.
A subset  $C $ of a partial order $(V, \le)$  is   $\le$-{\em convex}   if     for every $ u, v, z \in V $, if $ u, z \in C $ and $ u < v < z $, then $ v \in C $.
If $C$ is a  $\le$-convex set  over a total order, then we call it an $\le$-{\em interval}. 
If the order is  deducible from the context, we drop the prefix $\leq$.

If $ \Sigma $ is a finite alphabet of size $\sigma$, we denote by $ \Sigma^* $ the set of  (possibly empty) finite words over $\Sigma$.  We fix an order $\preceq $ over the alphabet $\Sigma$ and we extend it  co-lexicographically to words in $\Sigma^*$, that is, for $ \alpha, \beta \in \Sigma^* $ we declare $ \alpha \preceq \beta $ if and only if the reversed string $ \alpha^R $ is lexicographically smaller than or equal to $ \beta^R $. We call this order \emph{co-lex}, and in Section \ref{sec:colex orders} extend it to sets of strings and states of an NFA.

 A {\em nondeterministic finite automaton} (NFA) over the alphabet $ \Sigma $ is a $ 4 $-tuple $ \mathcal{A} = (Q, s, \delta, F) $, where $ Q $ is the set of states,   $s$ is the initial state, $\delta: Q\times \Sigma \rightarrow Pow(Q)$  is the transition function, and $ F \subseteq Q $ is the set of final states. As customary, we extend $ \delta $ to operate on strings as follows: for all $ q \in Q, a\in \Sigma, $ and $ \alpha \in \Sigma^{*} $:
\begin{align*}
 {\delta}(q,\epsilon)=\{q\}, &   \hspace{1cm}
 {\delta}(q,\alpha a) = \bigcup_{v\in{\delta}(q,\alpha)} \delta(v,a).
\end{align*}

It is also convenient to define an {\em edge} of 
  an automaton as a triple  $ (u, v, a) $ with $v\in \delta(u,a)$ and denote the set of edges of the automaton as $E_{\mathcal A}$---simply $E$ when $\mathcal A$ is clear from the context.
  
We say that a state $q'$ is {\em reachable} from a state $q$ if there exists $\alpha\in \Sigma^*$  with $q'\in \delta(q, \alpha)$. 
  We denote by   $\mathcal L(\mathcal A)=\{\alpha\in \Sigma^*\ |\  \delta(s, \alpha)\cap F\neq \emptyset\}$. An automaton  $ \mathcal A $  is {\em deterministic}   (a DFA),  if $ |\delta(q,a)|\leq 1 $,  for any $ q \in Q $ and $ a \in \Sigma $.  If the automaton is deterministic we  write  ${\delta}(q,\alpha)=q'$  for   the unique $q'$  such that ${\delta}(q,\alpha)=\{q'\}$  (if defined). 
  
  Throughout this paper we assume that every NFA $ \mathcal{A} $ satisfies the following properties: (1) every state is reachable from the initial state, (2) every state   allows to reach a final state,  (3) the initial state is not reachable from any other state, and (4) all edges reaching the same state have the same label (\emph{input-consistency}). This is required for indexing and is not restrictive since input-consistency can be forced by replacing each state with $ |\Sigma | $ copies of itself without changing the accepted language. To simplify our notation, we denote by $\lambda(u)$ the (uniquely determined) label of all incoming edges of node $u$. For the initial state $s$, we write $\lambda(v) = \# \notin \Sigma$ and we assume $\# \prec c$ for all $c\in\Sigma$. To make notation more compact, we will sometimes write $ (u, v) $ for $ (u, v, a) $, because it must be $ a = \lambda (v) $.

If $ \mathcal  A = (Q, s, \delta, F) $ is an NFA, we denote with     $ Pref (\mathcal{L(A)}) $ the set of   prefixes of   words in  $\mathcal L(A)$. For every $ \alpha \in Pref (\mathcal{L(A)}) $, let $ I_\alpha=\{q\in Q ~| ~ \delta(s, \alpha)=q\}$. For every $ q \in Q $, let $ I_q=\{\alpha\in  Pref (\mathcal{L(A)})~| ~\delta(s, \alpha)=q\} $.

\section{Co-lexicographic orders}\label{sec:colex orders}

In this section we recall basic definitions and results from \cite{cotumaccio2021indexing} and we explain how a co-lex order can mirror the co-lex order of the words in the sets $ I_u $'s. 

\begin{definition}\label{def:co-lex}
Let $\mathcal A = (Q, s, \delta, F) $ be an NFA. A \emph{co-lex order} on $ \mathcal A $ is a partial order $ \le $ on $ Q $ that satisfies the following two axioms:
\begin{enumerate}
    \item (Axiom 1) For every $ u, v \in Q $, if $\lambda(u) \prec \lambda(v)$, then $ u < v $ (in particular, the initial state comes before all remaining states);
    \item (Axiom 2) For all edges $ (u', u), (v', v) \in E $, if $ \lambda (u) = \lambda (v) $ and $ u < v $, then $ u' \leq v' $.
\end{enumerate}
\end{definition}

As originally defined in \cite{GAGIE201767}, \emph{Wheeler automata}  are precisely those for which 
the order $\le$  of Definition \ref{def:co-lex} is total. Not all automata admit a Wheeler order and not all languages are recognized by some Wheeler automaton. On the other hand, every automaton admits a co-lex order \cite{cotumaccio2021indexing}. As a consequence, if we drop the totality requirement, we can consider the whole class of finite automata   and we can use the \emph{width} of the partial order (intuitively, the ``distance''   from being a total order) to classify automata and the languages that they accept (see Definitions \ref{def:psrt_automata} and  \ref{def:psrt_languages}).

\begin{definition}\label{def:psrt_automata}
\cite[Def. 3.3]{cotumaccio2021indexing} Let $ \mathcal{A} = (Q, s, \delta, F) $ be an NFA. The width of  $ \mathcal{A} $, denoted by ${\text width}(\mathcal A) $,    is the smallest width of a co-lex order on $ \mathcal{A} $.
\end{definition}

Moving from states to set of strings, a co-lex order forces a partial order over the family of all $ I_q $'s. First, let us recall the key result from \cite{cotumaccio2021indexing} (the reader can find the original proof in the appendix).

\begin{lemma}\label{lem:string-nodes}
\cite[Lem. 3.1]{cotumaccio2021indexing} Let $ \mathcal{A} = (Q, s, \delta, F) $ be an NFA, and let $ \le $ be a co-lex order on $ \mathcal{A} $. Let $ u, v \in Q $ and $ \alpha, \beta, \in Pref (\mathcal{L(A)}) $ be such that $ u \in I_\alpha $, $ v \in I_\beta $ and $ \{u, v \} \not \subseteq I_\alpha \cap I_\beta $.
\begin{enumerate}
\item If $ \alpha \prec \beta $, then $ u ~\|~ v $ or $ u < v $.
\item If $ u < v $, then $ \alpha \prec \beta $.
\end{enumerate}
\end{lemma}

We can now define a partial order on the $ I_u $'s.

\begin{definition}
Let $ \mathcal{A} = (Q, s, \delta, F) $ be an NFA. On the set $ \{I_u\  |\ u \in Q \} $, define for $ I_u \not = I_v $:
\begin{equation*}
    I_u \prec I_v \iff (\forall \alpha \in I_u)(\forall \beta \in I_v)(\{\alpha, \beta\} \not \subseteq I_u \cap I_v \to \alpha \prec \beta).
\end{equation*}
\end{definition}

\begin{lemma}\label{lem:preceq}
Let $ \mathcal{A} = (Q, s, \delta, F) $ be an NFA. Then, $ (\{I_u | u \in Q \}, \preceq) $ is a partial order.
\end{lemma}

From Lemma \ref{lem:string-nodes} we immediately obtain:

\begin{corollary}\label{cor:setstrings}
Let $ \mathcal{A} = (Q, s, \delta, F) $ be an NFA. Let $ \le $ be a co-lex order on $ \mathcal{A} $. If $ u < v $, then $ I_u \preceq I_v $.
\end{corollary}

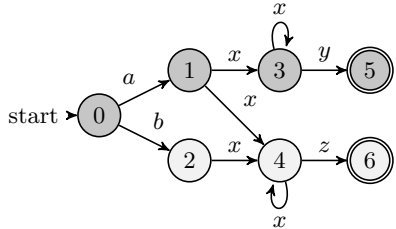
\begin{figure}[h!] 
 \begin{center}
\begin{tikzpicture}[->,>=stealth', semithick, auto, scale=.3]
\tikzset{every state/.style={minimum size=0pt}}

\node[state,fill=gray!45, initial] (s)    at (0,0)	{0 };

\node[state,fill=gray!45, label=above:{}] (a)    at (4,2)	{1 };
\node[state, fill=gray!10, label=above:{}] (b)    at (4,-2)	{ 2};
\node[state,fill=gray!45, label=above:{}] (x1)    at (8,2)	{3 };
\node[state,fill=gray!10,  label=above:{}] (x2)    at (8,-2)	{4 };
\node[state, fill=gray!45, accepting, label=above:{}] (c)    at (12,2)	{5 };
\node[state, fill=gray!10, accepting, label=above:{}] (d)    at (12,-2)	{6};

 \draw (s) edge  [] node [] {$a$} (a);
  \draw (s) edge [] node [] {$b$} (b);
 \draw (a) edge [] node [] {$x$} (x1);
 \draw (a) edge  [] node [] {$x$} (x2);
 \draw (b) edge  [] node [] {$x$} (x2);
 \draw (x1) edge  [] node [] {$y$} (c);
  \draw (x2) edge  [] node [] {$z$} (d);
  \draw (x1) edge  [loop above] node {$x$} (x1);
   \draw (x2) edge  [loop below] node {$x$} (x2);
\end{tikzpicture}
 \end{center}
 	\caption{An   NFA $ \mathcal{A} $ with ${\text width}(\mathcal A)=2$.}\label{fig:width2}
\end{figure}

In  Figure \ref{fig:width2} we present an NFA $ \mathcal{A} $  with ${\text width}(\mathcal A)=2$. The width is at most 2 because the reflexive and transitive closure of $ \{(0, 1), (1, 3), (3, 5), (0, 2), (2, 4), (4, 6) \} $ is a co-lex order. At the same time, the width cannot be 1, because (1) $ax,axx\in I_3$, (2) $bx\in I_{4}\setminus I_3$, (3) $ax\prec bx\prec axx$, so  using Corollary \ref{cor:setstrings} we conclude that states $3$ and $4$ cannot  be comparable in any co-lex order on $ \mathcal{A}$. 

In general, an NFA admits several co-lex orders. However, for DFAs we have the following result:

\begin{lemma}\label{lem:maxcolex}
Let $ \mathcal A= (Q, s, \delta, F) $ be a DFA. For every $ u, v \in Q $, with $ u \not = v $, let:
\begin{equation*}
   u < v \iff  I_u \prec I_v.
\end{equation*}
Then, $ \le $ is a co-lex order. Moreover, for every co-lex order $ \le' $ on $ \mathcal{A} $ and for every $ u, v \in Q $, if $ u \leq' v $, then $ u \leq v $. We say that $ \le $ is the \emph{maximum co-lex order} on $ \mathcal{A} $.
\end{lemma}

The existence of the maximum co-lex order has already been proved in \cite{cotumaccio2021indexing}, but here we provide an explicit characterization in terms of the $ I_u $'s that will be crucial for our results.

We end this section with a technical lemma from \cite{cotumaccio2021indexing} which will be used in Section \ref{sec:indexing}.
 For completeness, we provide the original proof in the appendix. 

\begin{lemma}\label{lem:transitive_closure}\cite[Lem. 6.1]{cotumaccio2021indexing}
Let $ \mathcal{A} = (Q, s, \delta, F) $ be an NFA, and let $ R $ be a relation   on $ Q $ such that:
\begin{enumerate}
    \item For every $ u, v \in Q $, if $\lambda(u) \prec \lambda(v)$, then $ (u, v) \in R $;
    \item For all edges $ (u', u), (v', v) \in E $, if $ \lambda (u) = \lambda (v) $ and $ (u, v) \in R $, then $ (u', v') \in R $.
\end{enumerate}
If  
  the transitive  and reflexive closure $ \mathcal R^* $ of $ \mathcal R $  is antisymmetric,  then $\mathcal R^* $ is a co-lex order on $ \mathcal{A} $.
\end{lemma}

 \section{Stratifying Regular Languages by \emph{Width}}\label{sec:width}

On the grounds of the definition of automata's co-lex width, we start studying its implications for the theory of regular languages. In this section, we
define the width of a regular language based on the co-lex orders on the automata recognizing it. As pointed out in \cite{cotumaccio2021indexing}, in Section \ref{sec:indexing} we shall see that the width of a language is a measure of algorithmic complexity in pattern matching.

\begin{definition}\label{def:psrt_languages}
Let $ \mathcal{L} $ be a regular language.
\begin{enumerate}
    \item The nondeterministic width   of $ \mathcal{L} $, denoted by ${\text width}^N(\mathcal L) $, is the smallest integer $ p $ for which there exists an NFA $ \mathcal{A} $ such that $ \mathcal{L(A)} = \mathcal{L} $ and ${\text width}(\mathcal A) = p $.
    \item The deterministic width   of $ \mathcal{L} $, denoted by ${\text width}^D(\mathcal L) $, is the smallest integer $ p $ for which there exists a DFA $ \mathcal{A} $ such that $ \mathcal{L(A)} = \mathcal{L} $ and ${\text width}(\mathcal A) = p $.
\end{enumerate}
\end{definition}

In Lemma \ref{lem:non-empty} we show that every level of both the above  hierarchies is non-empty. To this end, we shall use the following lemma:

\begin{lemma}\label{lem:NFAcycle}
Let $ \mathcal{A} = (Q, s, \delta, F) $ be an NFA. Assume that $ \mathcal{A} $ contains a simple cycle with $ m$ states such that all edges of the cycle are equally labeled. Then, ${\text width}(\mathcal A) \ge m $.
\end{lemma}

We can now prove:

\begin{lemma}\label{lem:non-empty}
For every integer $ p \ge 1 $, there exists a regular language $ \mathcal{L} $ such that $ {\text width}^N(\mathcal L) = {\text width}^D(\mathcal L) = p $.
\end{lemma}

Clearly, for every regular language $\mathcal L$ we have ${\text width}^N(\mathcal L)\leq {\text width}^D(\mathcal L)$.  Moreover, 
for languages  with ${\text width}^N(\mathcal L)=1$, the so-called {\em Wheeler languages}, it is known that 
the nondeterministic and deterministic width coincide \cite{alanko2020regular}. Nonetheless, we will prove that this property is truly peculiar of Wheeler languages, because the gap between the deterministic and nondeterministic hierarchies is, in general, exponential.

Let us start by providing an \emph{upper} bound. The idea is to transform an NFA into an equivalent DFA by the usual powerset construction. Recall that, given an NFA $ \mathcal{A} = (Q, s, \delta, F) $, the powerset construction algorithm builds a DFA $ \mathcal{A^*} = (Q^*, s^*, \delta^*, F^*) $ such that $ \mathcal{L(\mathcal{A})} = \mathcal{L(\mathcal{A^*})} $ defined as: 
(i) $ Q^* = \{ I_{\alpha}\ |\ \alpha\in Pref (\mathcal{L(\mathcal{A})})\}$, 
(ii) $ s^* = \{s \} $, (iii) $ \delta (I_\alpha, a) = I_{\alpha a} $ for all $ \alpha \in \Sigma^*$ and $ a \in \Sigma $ such that $ \alpha a \in Pref (\mathcal{L(A)}) $, and (iv) $ F^* = \{I_\alpha\ |\ \alpha \in \mathcal{L(\mathcal{A})} \} $.

\begin{lemma}\label{lem: 2^p}
Let $ \mathcal{A}$ be an NFA  and let  $ \mathcal{A^*}$ be the powerset automaton obtained from $ \mathcal{A} $. Then, ${\text width} (\mathcal{A^*}) \le 2^{{\text width} (\mathcal{A})} - 1 $.
\end{lemma}

By picking an NFA $ \mathcal{A} $ such that $ width (A) = {width}^N(\mathcal L) $, we immediately conclude:
 
\begin{corollary}\label{cor:widths}
Let $ \mathcal{L} $ be a regular language. Then, $ {\text width}^D(\mathcal L) \le 2^{{\text width}^N(\mathcal L)} - 1 $.
\end{corollary}

Notice that, when   ${\text width}^N(\mathcal L)=1$ (that is, when  $\mathcal L$ is a Wheeler language), we  obtain  that  ${\text width}^D(\mathcal L) ={\text width}^N(\mathcal L)$, as already proved in 
\cite{alanko2020regular}.



Next, the following lemma will imply an exponential \emph{lower} bound.

\begin{lemma}\label{lem:det_non_width}
Let $ p_1, \dots, p_k $ be distinct primes. Then, there exists a language $ \mathcal{L} $ such that $ {\text width}^D(\mathcal L) \ge \prod_{i = 1}^k p_i $ and $ {\text width}^N(\mathcal L) \le \sum_{i = 1}^k p_i $.
\end{lemma}

Our lower bound easily follows:

\begin{lemma}\label{lem:lower bound}
There exist a regular language $ \mathcal{L} $, whose width can be chosen arbitrarily large, such that $width^{D}(\mathcal L) \geq e^{\sqrt{width^{N}(\mathcal L)}}$.
\end{lemma}


Merging the above results with powerset construction and the indexing mechanism for DFAs described in \cite[Cor. 6.1]{cotumaccio2021indexing}, one can easily obtain that NFAs can be indexed in polynomial time while breaking the lower bound \cite{equi2020graphs} of  $\Omega(\pi m)$ for matching a pattern of length $\pi$ on the paths of an NFA of size $m$, whenever the NFA's width is $p < 0.5 \log_2 m$. In the next section we improve exponentially this indexability bound, raising it to $p \in o(\sqrt{m})$.

\section{Indexing NFAs in polynomial time}\label{sec:indexing}

In \cite{cotumaccio2021indexing} it was shown that, given a co-lex order of width $p$ for an NFA $\mathcal A$, it is possible to build an index for $\mathcal A$ solving pattern matching queries in $O(\patlen \cdot p^2\log(p\sigma))$ time, $\patlen $ being the pattern length. While the index can be built in polynomial time, finding the order yielding the smallest width $p$ (that is, the co-lex width of $\mathcal A$) is known to be an NP-hard problem \cite{cotumaccio2021indexing}.
In this section, we get around this problem by providing a polynomial-time algorithm that computes a new partial order for $\mathcal A$ (not necessarily a co-lex order) with the following interesting properties: (1) the new order's width is not larger than $\mathcal A$'s width, (2) while the new order is not necessarily co-lex, it satisfies a weaker property that we show to be sufficient for indexing. As a result, we will obtain a polynomial-time algorithm for indexing NFAs.

\subsection{Sorting NFAs}\label{sec:sorting}

In this subsection we present a polynomial-time algorithm computing a new indexable NFA order. Intuitively, our construction will work as follows:

\begin{enumerate}
    \item Algorithm \ref{alg:R(A,u,v)} solves the following problem: given an NFA $\mathcal A$ and two states $u,v$, decide whether there exists a co-lex order $\leq$ such that $u \leq v$. Denote by $(u,v) \in \rho_{\exists}$ such a property.
    \item Our final order $\trianglelefteq$ compares states inside the same strongly connected component (SCC) of $\rho_{\exists}$ arbitrarily (according to any pre-determined order over $\mathcal A$'s states), and states in distinct SCCs according to the reachability relation on $\rho_{\exists}$.
\end{enumerate}

Our main result is to show that the order $\trianglelefteq$, while not necessarily being co-lex, (i) has width no larger than $width(\mathcal A)$, (ii) can be computed in polynomial time (as opposed to the co-lex order of smallest width), and (iii) enables NFA indexing.

\begin{algorithm}[th!]
	\caption{$\mathtt{\mathcal R(\mathcal A, u,v)}$}\label{alg:R(A,u,v)}

	\SetKwInOut{Input}{input}
	\SetKwInOut{Output}{output}
	\SetSideCommentLeft
	\LinesNumbered
	
	\Input{An NFA $\mathcal A = (Q,s,\delta,F)$ and two distinct states $u,v\in Q,\ u\neq v$.}
	\Output{A relation $\rho \subseteq Q\times Q$ with the following properties: (1) any co-lex order containing $(u,v)$ must also contain the entire $\rho$, and (2) the reflexive and transitive closure of $\rho$ is a co-lex order containing $(u,v)$.\\ If no co-lex order containing $(u,v)$ exists, then it returns $\emptyset$.}
	\BlankLine

	$\rho \leftarrow \emptyset$\Comment*[r]{\small Initialize an empty relation} \label{init rel}
	
	$S \leftarrow \emptyset$ \Comment*[r]{\small Initialize an empty stack of state pairs yet to be processed} \label{init stack}

    $\rho \leftarrow \rho \cup \{(u,v)\}$\Comment*[r]{\small $\rho$ contains $(u,v)$ by definition} \label{insert uv}
    $S.push((u,v))$
    
    \BlankLine
    
    \ForEach{$u',v'\in Q$}{
    
        \If{$\lambda(u') \prec \lambda(v')$}{
    
            $\rho \leftarrow \rho \cup \{(u',v')\}$\Comment*[r]{\small Co-lex Axiom 1}\label{rho co-lex 1}
        
        }
    
    }
    
    \BlankLine
    
    \Repeat{$S = \emptyset$}{
    
        $(u',v') \leftarrow S.pop()$\;\label{pop}
        
        \ForEach{  $u'',v''\in Q\ | (\ u''\neq v''\ \wedge\ (\exists a\in\Sigma)( (u'',u',a),(v'',v',a) \in E_{\mathcal A} ))$ }{
        
        \If{$(v'',u'') \in \rho$}{ \Return $\emptyset$
        \Comment*[r]{\small $\rho$ is not antisymmetric}\label{no antisymmetric}}

        \If{$(u'' , v'') \notin \rho$}{ 
        
            $\rho \leftarrow \rho \cup \{(u'',v'') \}$\Comment*[r]{\small Co-lex Axiom 2}\label{rho co-lex 2}
            $S.push((u'',v''))$\;
        
        }

        }

    }

    \BlankLine

    \eIf{$acyclic(\rho)$}{\Return $\rho$\Comment*[r]{\small The transitive closure of $\rho$ is antisymmetric}\label{acyclic}
    }{\Return $\emptyset$\Comment*[r]{\small The transitive closure of $\rho$ is not antisymmetric}\label{cyclic}
    }

\end{algorithm}

\begin{lemma}\label{lem:nonemptyR}  
The output of Algorithm \ref{alg:R(A,u,v)} - that is, $ \mathcal R(\mathcal A,u,v) $ - is nonempty if and only if there exists a co-lex order  on $\mathcal A$ containing the pair $(u,v)$.  
\end{lemma}

\begin{lemma}\label{lem:complexity}
Algorithm \ref{alg:R(A,u,v)} can be implemented so that it terminates in $O(|Q|^4)$ time.
\end{lemma}

\begin{definition}\label{def:rhoexists}
Let $\rho_\exists$ be the relation on the set of states 
such that $ (u, v) \in \rho_\exists $ if and only if $ \mathcal R(\mathcal A, u,v) \not = \emptyset $.
Moreover, let $\sim$ be the equivalence relation such 
that $u \sim v$ if and only if $u$ and $v$ belong to 
the same strongly connected component of $\rho_\exists $.
\end{definition}

\begin{lemma}\label{lem:alg} Consider the relations $\rho_\exists$ and $\sim$.
\begin{enumerate}
    \item If $ (u, v) \in \rho_\exists $, then $ I_u \preceq I_v $;
    \item If $u \sim v$, then $ I_u = I_v $.
\end{enumerate}
\end{lemma}

Let $ \{q_1, \dots, q_n \} $ be an arbitrary fixed enumeration of the states of $ \mathcal{A} $.

\begin{definition}\label{def:triangle}
Let $ R\subseteq Q\times Q$ be the relation on the set of states of $ \mathcal{A} $ defined as follows. $(q_i,q_j)\in R$ if and only if:
\begin{itemize}
    \item $q_i \sim q_j$ and $i < j$, or
    \item $ q_i \not\sim q_j$ and $(q_i , q_j) \in \rho_\exists$.
\end{itemize}
Let, moreover, $ \trianglelefteq~\subseteq Q\times Q$ be the reflexive and transitive closure of $ R $.
\end{definition}

\begin{remark}\label{rem:path rho}
    If $(u,v) \in R$, then there exists a path from $u$ to $v$ in $\rho_\exists$. Indeed, If $u \sim v$, then by definition of $\sim$, the states $u$ and $v$ belong to the same strongly connected component of $\rho_\exists$ and we are done. On the other hand, if $u \not\sim v$ then by definition of $R$ we have $(u,v) \in \rho_\exists$.
\end{remark}

\begin{lemma}\label{lem: triangle properties}
The following properties hold:
\begin{enumerate}
    \item if $ u \trianglelefteq v $, then $ I_u \preceq I_v $;
    \item $ \trianglelefteq $ is a partial order;
    \item $ width (\trianglelefteq) \preceq width (\mathcal{A}) $;
    \item $ \trianglelefteq $ can be computed in $O(|Q|^6)$  time.
\end{enumerate}
\end{lemma}

Let $ \mathcal{A} $ be an NFA for which there exists the maximum co-lex order, that is, a co-lex order $ \le $ such that for every co-lex order $ \le' $ on $ \mathcal{A} $ we have that $ u \le' v $ implies $ u \le v $. Then, the following lemma shows  that $\trianglelefteq$ is equal to the maximum co-lex order. Since $\trianglelefteq$ can be determined in polynomial time, then the width of $ \mathcal{A} $ can be determined in polynomial time as well. This implies that the NP-hardness of determining the width of an NFA is due to automata that do not admit the maximum co-lex order. Lemma \ref{lem:maxcolex} implies that all DFAs admit the maximum co-lex order. More generally, it may be proved that all reduced NFAs (that is, NFAs such that   $u\neq v$ implies $I_u\neq I_v$) admit the maximum co-lex order.

\begin{lemma}\label{lem:equalrelation}
Let $ \mathcal{A} $ be an NFA that admits the maximum co-lex order $ \le $. Then, $\trianglelefteq$ and $ \le $ are the same relation.
\end{lemma}

\subsection{Indexing NFAs} \label{subsec:indexing}

In this section we introduce an index based on our NFA order $\trianglelefteq$.
In fact, our index works on any order satisfying a weaker property: \emph{weak path coherency} (Definition (\ref{def:wpc}) below).

Let $P \dashv S$, for $P,S\in \Sigma^*$, indicate that $P$ is a suffix of $S$.

\begin{definition}
 Let $\mathcal B(P) = \{u\in Q\ | (\exists \alpha\in I_u) (P \dashv \alpha)\}$ denote the set of all states reached by at least one path suffixed by $P$.
\end{definition}

\begin{definition}\label{def:wpc}
Let $\mathcal A = (Q,s,\delta,F)$ be an NFA, and $(Q,\leq)$ be a partial order. We say that $(Q,\leq)$ is \emph{weakly path coherent} if $\mathcal B(P)$ is a convex set in $(Q,\leq)$, for any $P\in \Sigma^*$.
\end{definition}

We first show that our order $\trianglelefteq$ is weakly path coherent, starting from the property in Lemma \ref{lem: triangle properties}-1:

 \begin{lemma}\label{lem:WPC}
 Let $\leq$ be any partial order over the states $Q$ of an NFA $\mathcal A = (Q,s,\delta,F)$, such that $u \leq v \Rightarrow I_u \preceq I_v$. Then, $(Q,\leq)$ is weakly path coherent.
 \end{lemma}

\begin{remark}
The name of the above property derives from the original work of \cite{GAGIE201767} introducing Wheeler graphs, where the \emph{path coherency} property was introduced. 
This original path coherency property (later generalized in \cite{cotumaccio2021indexing} to co-lex NFAs) states that, starting from \emph{any} interval in $(Q,\leq)$ and considering all states reachable by reading a fixed string $\alpha \in \Sigma^*$, we end up in another interval on $(Q,\leq)$. Weak path coherence is the special case in which the initial interval is the whole set $ Q $.

\end{remark}

In the following, let $(Q,\leq)$ be any weakly path coherent partial order. 
Fix a chain partition $Q_1, \dots, Q_t$ of $(Q,\leq)$ into $t$ linear (i.e., totally ordered) components. 
Note that a smallest chain decomposition of a partial order can be computed in polynomial time in the order's size (that is, number of comparable pairs) \cite{ford1962flows}.
In the following definition we consider the restriction of $\mathcal B(P)$ to the states of a single linear component:

\begin{definition}
$\mathcal B_i(P) = \mathcal B(P) \cap Q_i$, for each $i=1, \dots, t$.
\end{definition}

By Definition \ref{def:wpc}, the set of states $\mathcal B(P)$ can be expressed as a disjoint union of $t$ (possibly empty) \emph{intervals} on the $t$ linear components $Q_1, \dots, Q_t$, as follows: $\mathcal B(P) = \cup_{i=1}^t \mathcal B_i(P)$. In particular, each such interval can be encoded by two indices $l_i$ and $r_i$, as follows: $\mathcal B_i(P) = Q_i[l_i, r_i]$ ($l_i,r_i$ included), where $Q_i[k]$ denotes the $k$-th element in the total order $(Q_i, \leq)$. This means that $\mathcal B(P)$ can be compactly represented as the $t$-uple of pairs $(l_i,r_i)_{i=1}^t$. Indeed, in \cite{cotumaccio2021indexing} it was shown that such a representation is asymptotically optimal, for $t = width(\leq)$ (that is, $\Omega(width(\leq))$ words are required in the worst case to represent $\mathcal B(P)$). 

The idea behind our index is the following. For simplicity, we first give a description of our structure using arrays,  then turn the arrays into  data structures supporting fast queries. 
Let $OUT_i[1,|Q_i|]$, for $i=1,\dots, t$, be an array such that $OUT_i[k]$ contains all outgoing edges of the $k$-th node in the $i$-th linear component, in the following format:  $OUT_i[k] = \langle (a,j,q)\ :\ (Q_i[k], Q_j[q], a) \in E_{\mathcal A} \rangle$. In other words, for each node we build a list (sorted arbitrarily) of all its outgoing edges, storing, for each of them, the edge's label $a$, the target linear component $Q_j$, and the index $q$ of the target node $Q_j[q]$.

We argue that arrays $OUT_i$, for $i=1, \dots, t$, are sufficient for solving pattern matching queries. Let $(l_i,r_i)_{i=1}^t$ be the intervals, on the $t$ linear components, of all nodes reached by a path suffixed by $P$, i.e. the representation of $\mathcal B(P)$. 
Letting $a\in \Sigma$,
we show how to update those intervals and compute $(l'_i,r'_i)_{i=1}^t$:  the intervals of all nodes reached by a path suffixed by $Pa$, i.e. $\mathcal B(Pa)$.
By the weak path coherency property (Definition \ref{def:wpc}), each $B_j(Pa) = Q_j[l'_j,r'_j]$ is an interval; in particular, we can compute this interval by retrieving just its minimum $l'_j$ and maximum $r'_j$. This is easy to achieve by means of the following technique: iterate over all origin intervals $Q_i[l_i,r_i]$ (states reached by pattern $P$), keeping track (scanning $OUT_i[l_i,r_i]$) of the minimum and maximum index of a node in $Q_j$ reached by an edge labeled with character $a$ and originating from $Q_i[l_i,r_i]$. The following algorithm formalizes this procedure.

\begin{algorithm}[th!]
	\caption{Forward extension}\label{alg:extension}

	\SetKwInOut{Input}{input}
	\SetKwInOut{Output}{output}
	\SetSideCommentLeft
	\LinesNumbered
	
	\Input{The convex set $\mathcal B(P)$, expressed as $t$ intervals $(l_i,r_i)_{i=1}^t$, and a letter $a\in \Sigma$.}
	\Output{The convex set $\mathcal B(Pa)$, expressed as $t$ intervals $(l'_i,r'_i)_{i=1}^t$.}
	\BlankLine

	 \ForEach{$j = 1,\dots, t$}{
	 
	    $m \leftarrow |Q_j|+1$\;
	    $M \leftarrow 0$\;
	    
	    	 \ForEach{$i = 1,\dots, t$}{
            
                \ForEach{$s = l_i, \dots, r_i$}{
            
                        \ForEach{$(a',j',q) \in OUT_i[s]\ |\ a'=a \wedge j'=j$}{
            
                            $m \leftarrow \min(m,q)$\;
                            $M \leftarrow \max(M,q)$\;

                          }
            
                }
            
            }
	 
	        $( l'_j,r'_j ) \leftarrow ( m,M )$\;
	 
	 }

    \Return $(l'_i,r'_i)_{i=1}^t$\;

\end{algorithm} 

We now show how to speed up Algorithm \ref{alg:extension}.
We leave unchanged the two outer \texttt{for} loops  ($t^2$ iterations), and optimize searching the minimum and maximum elements in arrays $OUT_i$. Indeed,  plugging standard compact data structures  we can replace the two inner \texttt{for} loops with operations taking just $O(\log(t\cdot \sigma))$ time. In the appendix, we show: 

\begin{lemma}\label{lem:DS}
    There is a representation for arrays $OUT_i$ taking $O(|\mathcal A|)$ words of space and  simulating the two inner \texttt{for} loops of Algorithm \ref{alg:extension} in $O(\log(t\cdot \sigma))$ time.
\end{lemma}
\begin{proof}
\emph{(sketch)} We represent each $OUT_i$ as several sequences $W_{i,a,j} = \langle q\ |\ (Q_i[k], Q_{j}[q],a) \in E_{\mathcal A}  \rangle$ storing the positions $q$ of nodes in the $j$-th linear component reached from $Q_i$ while following edges labeled with character $a$. The particular representation we use (wavelet trees \cite{grossi2003high,ferragina2009myriad}), employs a \emph{cascading} technique allowing, given any range $l, r$, to find the sub-sequence  $W_{i,a,j}[l',r'] = \langle q\ |\ (Q_i[k], Q_{j}[q],a) \in E_{\mathcal A}\ \wedge \ l \leq k \leq r \rangle$. To conclude, a range minimum (maximum) data structure \cite{fischer2010optimal} allows finding the minimum (maximum) element in any range of $W_{i,a,j}$ in constant time.  \qed
\end{proof}

We finally obtain: 

\begin{theorem}\label{th:FM-index}
Let $ \mathcal{A} = (Q, s, \delta, F) $ be an NFA with $p = width(\mathcal A)$. In polynomial time we can build a data structure taking $O(|\mathcal{A}|)$ words of space that, given a query string $\alpha\in \Sigma^\patlen$,  supports the following operations in $O(\patlen\cdot p^2 \cdot \log(p\cdot \sigma))$ time:  
\begin{itemize}
    \item[(i)] Count the number of states reached by a path labeled $\alpha$.
    \item[(ii)] Return unique identifiers for the states reached by a path labeled $\alpha$. 
    \item[(iii)] Decide whether $\alpha \in \mathcal L(\mathcal A)$.
\end{itemize}
\end{theorem}

Notably, Theorem \ref{th:FM-index} allows to index in polynomial time the well-studied case $p=1$: Wheeler graphs \cite{GAGIE201767}. In more general terms, it shows that problems involving pattern matching on NFAs (for example, regular expression matching) can be solved quickly when the NFA's width is small.

\section{Conclusions and further developments}

We studied the collection of partial orders on the set of states of a given automaton $\mathcal A$ that maintain some coherence with the collection of strings reaching $\mathcal A$'s states.
We defined $\mathcal A$'s width as the minimum possible width of a co-lex order on the automaton and introduced two (the deterministic and the non-deterministic) non collapsing hierarchies of regular languages. We also showed that the  levels of such hierarchies are meaningful and proper complexity measures.
Although the first level of each of the two hierarchies captures the same class of languages---the so-called Wheeler languages, where we can also find unique minimal automata up to isomorphism \cite{alanko2020regular}---we proved that this is no longer true for higher levels, where we have an exponential gap between the nondeterministic and the deterministic hierarchy. This motivates the problem of indexing NFAs, which we showed can be solved in polynomial time. 

Our language-theoretic and algorithmic results find important applications to the study of regular expression matching algorithms: we showed that regular languages represented as NFAs $\mathcal A$ in the low ($p\in o(\sqrt{|\mathcal A|})$) levels of the nondeterministic hierarchy admit indexes supporting fast pattern matching queries (in particular, breaking known lower bounds) that can be built in polynomial time. 

In a paper in preparation we shall consider the following questions on the notion of width of an automaton.

\begin{enumerate}
    \item Given a regular language $\mathcal L$ (say, by giving its minimum DFA) can we calculate its width in 
    an effective way? Notice that the width of the minimum DFA does not, in general,  reflect the width of the language  already at level one: there are Wheeler languages for which the minimum  DFA is not Wheeler (see \cite{alanko2020regular}), so 
    the question is not trivial.
   \item As for other interesting subclasses of regular languages, Wheeler languages admit an automata free
   characterization: a language $\mathcal L$ is Wheeler if and only if 
  every monotone sequence in $(Pref(\mathcal L), \preceq)$ is  ``thin'', i.e. it   ends definitely  in at most one  Myhill-Nerode class \cite{alanko2020regular}. Can we find a similar characterization for 
 languages of width $p$, for $p>1$?
 \item Is it possible to derive the width of a language directly from some combinatorial/graph-theoretical property of the minimum DFA accepting $\mathcal L$? 
    \item Our lower bound of Lemma \ref{lem:lower bound} and upper bound of Lemma \ref{lem: 2^p} do not match. Can we improve this result by providing \emph{tight} bounds for the separation between the deterministic and nondeterministic hierarchies?  
\end{enumerate}

In addition, our work opens further intriguing questions of more algorithmic flavor. For instance, can we devise  fast algorithms that, given a DFA/NFA, outputs an equivalent DFA/NFA of minimum width?  Can we prove conditional lower bounds for the regular expression matching problem as a function of the language's width?

Notice that Algorithm \ref{alg:R(A,u,v)} can be implemented in polynomial time, but the running time is quite high (Lemma \ref{lem:complexity}). In this paper we focused on how to give an easy description of a polynomial time indexing technique for arbitrary NFAs, so overcoming the NP-hardness results for co-lex orders, but we did not explore efficient techinques for reducing the time complexity. For example, if $\mathcal R(\mathcal A,u,v) \neq \emptyset$, then we have basically built a co-lex order $ \le $, and we can immediately conclude that it also holds $\mathcal R(\mathcal A,u',v') \neq \emptyset$ for every $ u' $, $ v' $ such that $ u' < v' $ without needing to run Algorithm \ref{alg:R(A,u,v)} on the pair $ (u', v') $. 

Finally, can we adapt the circle of ideas and techniques developed in this area to other, more expressive, kind of networks?

\appendix

\section{Proofs of Section \ref{sec:colex orders}}

\paragraph*{\textbf{Statement of Lemma \ref{lem:string-nodes}.}} Let $ \mathcal{A} = (Q, s, \delta, F) $ be an NFA, and let $ \le $ be a co-lex order on $ \mathcal{A} $. Let $ u, v \in Q $ and $ \alpha, \beta, \in Pref (\mathcal{L(A)}) $ be such that $ u \in I_\alpha $, $ v \in I_\beta $ and $ \{u, v \} \not \subseteq I_\alpha \cap I_\beta $.
\begin{enumerate}
\item If $ \alpha \prec \beta $, then $ u ~\|~ v $ or $ u < v $.
\item If $ u < v $, then $ \alpha \prec \beta $.
\end{enumerate}

\paragraph*{\textbf{Proof}}
Since $ \{u, v\} \not \subseteq I_\alpha \cap I_\beta $, then either $ u \in I_\alpha \setminus I_\beta $ or $ v \in I_\beta \setminus I_\alpha $. Hence $ \alpha \not = \beta $ and $ u \not = v $.
\begin{enumerate}
\item We proceed by induction on $ \min( |\alpha|, | \beta|) $. If $ \min( |\alpha|, | \beta|) = 0 $, then $ \alpha = \epsilon $, so $ u = s $. We conclude $ u = s < v $ by Axiom 1.

Now assume $ \min( |\alpha|, | \beta|) \ge 1 $. This implies $ \alpha \not = \epsilon \not = \beta $ and $ u \not = s \not = v $. Let $ a $ be the last letter of $ \alpha $ and let $ b $ the last letter of $ \beta $; it must be $ a \preceq b $. If $ a \prec b $, then $ \lambda(u) \prec \lambda(v) $, which implies $ u < v $ by Axiom 1. Otherwise, we can write $ \alpha = \alpha' e $ and $ \beta = \beta' e $, with $ e \in \Sigma $, $ \alpha', \beta' \in \Sigma^* $ and $ \alpha' \prec \beta' $. Let $ u', v' \in  Q $ be such that $ u' \in I_{\alpha'} $, $ v' \in I_{\beta'} $ and there exist edges $ (u', u) $ and $ (v', v) $. Then $ \{u', v' \} \not \subseteq I_{\alpha'} \cap I_{\beta'} $, otherwise $ \{u, v \} \subseteq I_\alpha \cap I_\beta $. By the inductive hypothesis, we have $ u' ~\|~ v' $ or $ u' < v' $. Hence it must be $ u ~\|~ v $ or $ u < v $, otherwise it would be $ v < u $, which implies $ v' \le u' $ by Axiom 2.

\item We know that $ \alpha \not = \beta $. If it were $ \beta \prec \alpha $, then by the previous part it would be $ v ~\|~ u $ or $ v < u $, leading to a contradiction.
  \hfill $\Box$
\end{enumerate}

\paragraph*{\textbf{Statement of Lemma \ref{lem:preceq}.}} Let $ \mathcal{A} = (Q, s, \delta, F) $ be an NFA. Then, $ (\{I_u | u \in Q \}, \preceq) $ is a partial order.

\paragraph*{\textbf{Proof}}
    Let us prove antisymmetry. Assume that $ I_u \prec I_v $. Let us prove that $ I_v \not \prec I_u $. In particular, we have $ I_u \not = I_v $, so there exists $ \alpha \in I_u \setminus I_v $ or $ \beta \in I_v \setminus I_u $. Assume that there exists $ \alpha \in I_u \setminus I_v $ (the other case is analogous). It must be $ I_v \not = \emptyset $, so pick $ \beta \in I_v $. We have $ \{\alpha, \beta \} \not \subseteq I_u \cap I_v $, so $ I_u \prec I_v $ implies $ \alpha \prec \beta $. This means that $ I_v \prec I_u $ cannot hold, otherwise it should also be $ \beta \prec \alpha $, a contradiction.
    
    Let us prove transitivity. Assume that $ I_u \prec I_v $ and $ I_v \prec I_z $. Let us prove that $ I_u \prec I_z $. Pick $ \alpha \in I_u $ and $ \gamma \in I_z $ such that $ \{\alpha, \gamma \} \not \subseteq I_u \cap I_z $. We must prove that $ \alpha \prec \gamma $. Assume that $ \alpha \in I_u \setminus I_z $ (the other case, $ \gamma \in I_z \setminus I_u $, is analogous). We distinguish two cases.
    \begin{enumerate}
        \item Assume $ \alpha \in I_v $. Then $ \alpha \in I_v \setminus I_z $, so $ \{\alpha, \gamma \} \not \subseteq I_v \cap I_z $. From $ I_v \prec I_z $ it follows $ \alpha \prec \gamma $.
        \item Assume $ \alpha \not \in I_v $. We distinguish two subcases.
        \begin{enumerate}
            \item Assume $ \gamma \in I_v $. Then $ \alpha \in I_u \setminus I_v $ and $ \gamma \in I_v $, so $ I_u \prec I_v $ implies $ \alpha \prec \gamma $.
            \item Assume $ \gamma \not \in I_v $. Since $ I_v \not = \emptyset $, pick any $ \beta \in I_v $. We have $ \alpha \in I_u \setminus I_v $ and $ \beta \in I_v $, so $ I_u \prec I_v $ implies $ \alpha \prec \beta $. Moreover, we have $ \gamma \in I_z \setminus I_v $ and $ \beta \in I_v $, so $ I_v \prec I_z $ implies $ \beta \prec \gamma $. From $ \alpha \prec \beta $ and $ \beta \prec \gamma $ we conclude $ \alpha \prec \gamma $.
        \end{enumerate}
    \end{enumerate}
  \hfill $\Box$

\paragraph*{\textbf{Statement of Lemma \ref{lem:maxcolex}.}} Let $ \mathcal A= (Q, s, \delta, F) $ be a DFA. For every $ u, v \in Q $, with $ u \not = v $, let:
\begin{equation*}
   u < v \iff  I_u \prec I_v.
\end{equation*}
Then, $ \le $ is a co-lex order. Moreover, for every co-lex order $ \le' $ on $ \mathcal{A} $ and for every $ u, v \in Q $, if $ u \leq' v $, then $ u \leq v $. We say that $ \le $ is the \emph{maximum co-lex order} on $ \mathcal{A} $.

\paragraph*{\textbf{Proof}}
First, $ \le $ is a partial order by Lemma \ref{lem:preceq}. Let us prove Axiom 1. If for some $ u, v \in Q $ it holds $ \lambda (u) \prec \lambda (v) $, then every string in $ I_u $ ends with $ \lambda (u) $ and every strings in $ I_v $ ends with $ \lambda (v) $, hence we conclude $ u < v $ (in particular, this works for $ u = s $ also). Let us prove Axiom 2. Consider two edges $ (u', u), (v', v) \in E $ such that $ \lambda (u) = \lambda (v) $ and $ u < v $. We want to prove that $ u' < v' $. Fix $ \alpha' \in I_{u'} $ and $ \beta' \in I_{v'} $; we must prove that $ \alpha' \prec \beta' $. Let $ c = \lambda (u) = \lambda (v) $. We have $ \alpha' c \in I_u $ and $ \beta' c \in I_v $, so from $ u < v $ it follows $ \alpha' c \prec \beta' c $ and so $ \alpha' \prec \beta' $.

Finally, let us prove that $ \le $ is the maximum co-lex order. Let $ \le' $ be a co-lex order on $ \mathcal{A} $, and assume that $ u <' v $; we must prove that  $ u < v $. Fix $ \alpha \in I_u $ and $ \beta \in I_v $; we must prove that $ \alpha \prec \beta $. Since $ I_u \cap I_v = \emptyset $ (being $ \mathcal{A} $ a DFA), the conclusion follows from Lemma \ref{lem:string-nodes}.
  \hfill $\Box$

\paragraph*{\textbf{Statement of Lemma \ref{lem:transitive_closure}.}} Let $ \mathcal{A} = (Q, s, \delta, F) $ be an NFA, and let $ R $ be a relation   on $ Q $ such that:
\begin{enumerate}
    \item For every $ u, v \in Q $, if $\lambda(u) \prec \lambda(v)$, then $ (u, v) \in R $;
    \item For all edges $ (u', u), (v', v) \in E $, if $ \lambda (u) = \lambda (v) $ and $ (u, v) \in R $, then $ (u', v') \in R $.
\end{enumerate}
If the transitive  and reflexive closure $ \mathcal R^* $ of $ \mathcal R $  is antisymmetric,  then $\mathcal R^* $ is a co-lex order on $ \mathcal{A} $.

\paragraph*{\textbf{Proof}}
Clearly, $ \mathcal R^* $ is a partial order. Morover, $\mathcal R^* $ satisfies Axiom 1 of co-lex orders, because if $ u, v \in Q $ are such that $ \lambda (u) \prec \lambda (v) $, then $ (u, v) \in R $ and so $ (u, v) \in R^* $. So we just have to prove that Axiom 2 is satisfied. Consider two edges $ (u', u), (v', v) \in E $ such that $ \lambda (u) = \lambda (v) $ and $ (u, v) \in R^* $; we must prove that $ (u', v') \in R^* $. Since $ \mathcal R^* $ is the transitive and reflexive  closure of $\mathcal R $, there exist states $ z_1, \dots, z_r $ ($ r \ge 0 $) such that $ (u, z_1) \in R $, $ (z_1, z_ 2) \in R $, $ \dots $, $ (z_r, v) \in R $, and in particular $ (u, z_1) \in R^* $, $ (z_1, z_ 2) \in R^* $, $ \dots $, $ (z_r, v) \in R^* $. Since $ \lambda (u) = \lambda (v) $, then $ \lambda (u) = \lambda (z_1) = \dots = \lambda (z_r) = \lambda (v) $. Indeed, if for some $ j $ it were for example $ \lambda (z_j) \succ \lambda (u) = \lambda (v) $, then by Axiom 1 it should be $ (v, z_j) \in R $ and so $ (v, z_j) \in R^* $, which contradicts $ (z_j, v) \in R^* $, since $z_j\neq v$ and $R^*$ is antisymmetric. In particular, since $ u $ and $ v $ have incoming edges, then   all $ z_i$'s have incoming edges $ (z'_i, z_i) \in E $, for $ i = 1, \dots, k $. The second assumption implies that $ (u', z'_1) \in R $, $ (z'_1, z'_2) \in R $, $ \dots $, $ (z'_k, v') \in R $, so $ (u', z'_1) \in R^* $, $ (z'_1, z'_2) \in R^* $, $ \dots $, $ (z'_k, v') \in R^* $ and we conclude $ (u', z') \in R^* $.
  \hfill $\Box$

\section{Proofs of Section \ref{sec:width}}

\paragraph*{\textbf{Statement of Lemma \ref{lem:NFAcycle}.}} Let $ \mathcal{A} = (Q, s, \delta, F) $ be an NFA. Assume that $ \mathcal{A} $ contains a simple cycle with $ m$ states such that all edges of the cycle are equally labeled. Then, ${\text width}(\mathcal A) \ge m $.

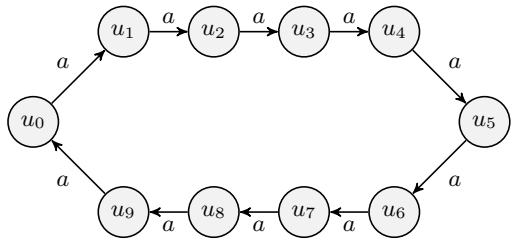
\begin{figure}[h!] 
 \begin{center}
\begin{tikzpicture}[->,>=stealth', semithick, auto, scale=.3]
\tikzset{every state/.style={minimum size=0pt}}

\node[state,fill=gray!10, ] (0)    at (0,0)	{$ u_0 $ };

\node[state,fill=gray!10, label=above:{}] (1)    at (4,4)	{$ u_1 $ };
\node[state, fill=gray!10, label=above:{}] (2)    at (8,4)	{$ u_2 $};
\node[state,fill=gray!10, label=above:{}] (3)    at (12,4)	{$ u_3 $ };
\node[state,fill=gray!10,  label=above:{}] (4)    at (16,4)	{$ u_4 $ };
\node[state, fill=gray!10, label=above:{}] (5)    at (20,0)	{$ u_5 $ };
\node[state, fill=gray!10, label=above:{}] (6)    at (16,-4)	{$ u_6 $};
\node[state, fill=gray!10, label=above:{}] (7)    at (12,-4)	{$ u_7 $};
\node[state, fill=gray!10, label=above:{}] (8)    at (8,-4)	{$ u_8 $};
\node[state, fill=gray!10, label=above:{}] (9)    at (4,-4)	{$ u_9 $};

 \draw (0) edge  [] node [] {$a$} (1);
  \draw (1) edge [] node [] {$a$} (2);
 \draw (2) edge [] node [] {$a$} (3);
 \draw (3) edge  [] node [] {$a$} (4);
 \draw (4) edge  [] node [] {$a$} (5);
 \draw (5) edge  [] node [] {$a$} (6);
  \draw (6) edge  [] node [] {$a$} (7);
  \draw (7) edge  [] node {$a$} (8);
   \draw (8) edge  [] node {$a$} (9);
      \draw (9) edge  [] node {$a$} (0);
\end{tikzpicture}
 \end{center}
 	\caption{No pairs of states in a cycle can be comparable.}\label{fig:cycle}
\end{figure}

\paragraph*{\textbf{Proof}}
The idea is that it must be ${\text width}(\mathcal A) \ge m $ because no pair of distinct states in the simple cycle can be comparable in any co-lex order. For simplicity, we provide an example that can be straightforwardly generalized to obtain a complete proof. Consider the cycle in Figure \ref{fig:cycle}, and suppose for sake of contradiction that there exists a co-lex order $ \le $ such that - say - it holds $ u_8 < u_5 $. Then, Axiom 2 of co-lex order implies that it must be $ u_7 < u_4 $, and then $ u_6 < u_3 $ and $ u_5 < u_2 $. To sum up, we have $ u_8 < u_5 < u_2 $. Proceeding again backward from $ u_5 < u_2 $, we obtain $ u_2 < u_9 $, so $ u_8 < u_5 < u_2 < u_9 $. Iterating this argument, we conclude $ u_8 < u_5 < u_2 < u_9 < u_6 < u_3 < u_0 < u_7 < u_4 < u_1 < u_8 $. In particular, $ u_8 < u_8 $, a contradiction. In general, since the cycle is simple we can always proceed backward without ending in the same state, and at some point the same state most occur twice in the chain of inequalities, leading to a contradiction.

 \hfill $\Box$

\paragraph*{\textbf{Statement of Lemma \ref{lem:non-empty}.}} For every integer $ p \ge 1 $, there exists a regular language $ \mathcal{L} $ such that $ {\text width}^N(\mathcal L) = {\text width}^D(\mathcal L) = p $.

\paragraph*{\textbf{Proof}}
Define:
\begin{equation*}
    \mathcal{L}_p  = \{a^{kp} \mid k \ge 0 \}.
\end{equation*}

\begin{figure}[h!] 
 \begin{center}
\begin{tikzpicture}[->,>=stealth', semithick, auto, scale=.3]
\tikzset{every state/.style={minimum size=0pt}}

\node[state,fill=gray!10, initial, accepting ] (0)    at (0,0)	{$ u_0 $ };

\node[state,fill=gray!10, label=above:{}] (1)    at (4,4)	{$ u_1 $ };
\node[state, fill=gray!10, label=above:{}] (2)    at (8,4)	{$ u_2 $};
\node[state, fill=gray!10, label=above:{}] (3)    at (12,0)	{$ u_3 $ };
\node[state, fill=gray!10, label=above:{}] (4)    at (8,-4)	{$ \dots $};
\node[state, fill=gray!10, label=above:{}, accepting] (5)    at (4,-4)	{$ u_p $};

 \draw (0) edge  [] node [] {$a$} (1);
  \draw (1) edge [] node [] {$a$} (2);
 \draw (2) edge [] node [] {$a$} (3);
 \draw (3) edge  [dashed] node [] {$a$} (4);
 \draw (4) edge  [dashed] node [] {$a$} (5);
 \draw (5) edge  [] node [] {$a$} (1);

\end{tikzpicture}
 \end{center}
 	\caption{A DFA recognizing $ \mathcal{L}_p $.}\label{fig:detwidth}
\end{figure}
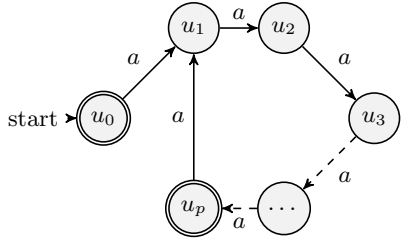

It will suffice to prove that $ \mathcal{L}_p $ is recognized by some DFA of width at most  $ p $  but it cannot be recognized by any  NFA of width at most $p-1$. First, The DFA in Figure \ref{fig:detwidth} recognizes $ \mathcal{L}_p $ and the maximum co-lex order on it has width at most $ p $ (because the DFA has $ p + 1 $ states and the initial state is comparable with any state).

Now, consider any NFA $ \mathcal{A}$ that recognizes $ \mathcal{L}_p $, whose alphabet must be $ \Sigma = \{a \} $.  Since    $\mathcal{L}_p$ is an infinite language, then the NFA  must  contain  a simple cycle $\mathcal{C}$. Let $ c $ be the length of the cycle $ \mathcal{C} $. Let $ u $ be any node in $ \mathcal{C} $ and let $ h $ and $ k $ be the lengths (number of edges) of some paths from $ s $ to $ u $ and from $ u $ to a final state, respectively. Then, it must $ a^{h + k }, a^{h + c + k} \in \mathcal{L}_p $, which implies that $c$ is a non-zero multiple of $p $. By Lemma \ref{lem:NFAcycle} we conclude that the width of $ \mathcal{A} $ cannot be smaller than $ p $.
  \hfill $\Box$

\paragraph*{\textbf{Statement of Lemma \ref{lem: 2^p}.}}
Let $ \mathcal{A}$ be an NFA  and let  $ \mathcal{A^*}$ be the powerset automaton obtained from $ \mathcal{A} $. Then, ${\text width} (\mathcal{A^*}) \le 2^{{\text width} (\mathcal{A})} - 1 $.

\paragraph*{\textbf{Proof}}
Let $ \le $ be a co-lex order on $ \mathcal{A} $ such that $ width (\le) = width (\mathcal{A}) $, and let $ \{Q_i \}_{i = 1}^{i = width (\mathcal{A})} $ be a $ \le $-chain partition. Let $ \le^* $ be the maximum co-lex order on $ \mathcal{A} $. Notice that the definition of the powerset automaton implies that:
\begin{equation*}
    I_\alpha <^* I_\beta \iff (\forall \alpha', \beta' \in Pref (\mathcal{L(A)}))((I_{\alpha'} = I_{\alpha}) \land (I_{\beta'} = I_\beta) \to \alpha' \prec \beta')
\end{equation*}
For every nonempty $ K \subseteq \{1, \dots, width (\mathcal{A}) \} $, define:
\begin{equation*}
    \mathcal{I}_K = \{I_\alpha \mid(\forall i \in \{1, \dots, k \})(I_\alpha \cap Q_i \not = \emptyset \iff i \in K) \}.
\end{equation*}
Notice that every $ I_\alpha $ belongs to exactly one $ \mathcal{I}_K $, so $ \{\mathcal{I}_K ~| ~\emptyset \subsetneqq K \subseteq \{1, \dots, width (\mathcal{A}) \}, \mathcal{I}_K \not = \emptyset \} $ is a partition of the set of states of $ \mathcal{A^*}$ having cardinality at most $ 2^{{\text width} (\mathcal{A})} - 1 $. As a consequence, it will suffice to prove that each $ \mathcal{I}_k $ is a $ \le^* $-chain. Fix $ I_\alpha, I_\beta \in \mathcal{I}_K $, with $ I_\alpha \not = I_\beta $. We must prove that $ I_\alpha $ and $ I_\beta $ are $ \le^* $-comparable. Since $ I_\alpha \not = I_\beta $, there exists either $ u \in I_\alpha \setminus I_\beta $ or $ v \in I_\beta \setminus I_\alpha $. Assume that there exists $ u \in I_\alpha \setminus I_\beta $ (the other case is analogous). In particular, let $ i \in \{1, \dots, width (\mathcal{A}) \} $ be the unique integer such that $ u \in Q_i $. Since $ I_\alpha, I_\beta \in \mathcal{I}_K $, from the definition of $ \mathcal{I}_K $ it follows that there exists $ v \in I_\beta \cap Q_i $. Notice that $ \{u, v \} \not \subseteq I_\alpha \cap I_\beta $ (so in particular $ u \not = v $), and since $ u, v \in Q_i $ we conclude that $ u $ and $ v $ are $ \le $-comparable. If $ u < v $, we conclude $ I_\alpha <^* I_\beta $ by Lemma \ref{lem:string-nodes}, because if $ \alpha', \beta' \in Pref (\mathcal{L(A)}) $ are such that $ I_{\alpha'} = I_\alpha $ and $ I_{\beta'} = I_\beta $, then $ u \in I_{\alpha'} $, $ v \in I_{\beta'} $, $ \{u, v \} \not \subseteq I_{\alpha'} \cap I_{\beta'} $ and $ u < v $, so we conclude $ \alpha' \prec \beta' $. Analogously, if $ v < u $, we conclude $ I_\beta <^* I_\alpha $. In both cases, $ I_\alpha $ and $ I_\beta $ are $ \le^* $-comparable.
 \hfill $\Box$

\paragraph*{\textbf{Statement of Lemma \ref{lem:det_non_width}.}} Let $ p_1, \dots, p_k $ be distinct primes. Then, there exists a language $ \mathcal{L} $ such that $ {\text width}^D(\mathcal L) \ge \prod_{i = 1}^k p_i $ and $ {\text width}^N(\mathcal L) \le \sum_{i = 1}^k p_i $.

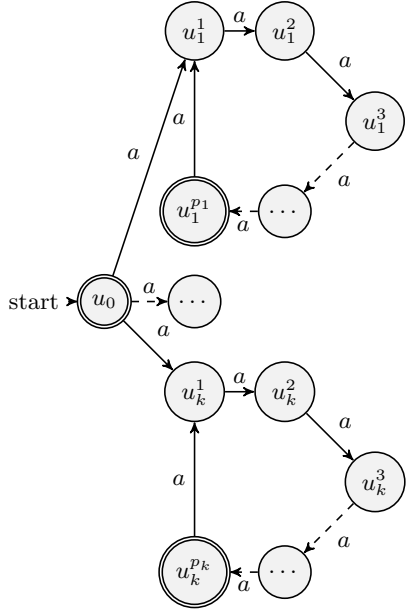
\begin{figure}[h!] 
 \begin{center}
\begin{tikzpicture}[->,>=stealth', semithick, auto, scale=.3]
\tikzset{every state/.style={minimum size=0pt}}

\node[state,fill=gray!10, initial, accepting ] (0)    at (0,-8)	{$ u_0 $ };

\node[state,fill=gray!10, label=above:{}] (1)    at (4,4)	{$ u_1^{1} $ };
\node[state, fill=gray!10, label=above:{}] (2)    at (8,4)	{$ u_1^{2} $};
\node[state, fill=gray!10, label=above:{}] (3)    at (12,0)	{$ u_1^{3} $ };
\node[state, fill=gray!10, label=above:{}] (4)    at (8,-4)	{$ \dots $};
\node[state, fill=gray!10, label=above:{}, accepting] (5)    at (4,-4)	{$ u_1^{p_1} $};
\node[state, fill=gray!10, label=above:{}] (6)    at (4, -8)	{$ \dots $ };
\node[state,fill=gray!10, label=above:{}] (7)    at (4,-12)	{$ u_k^1 $ };
\node[state, fill=gray!10, label=above:{}] (8)    at (8,-12)	{$ u_k^2 $};
\node[state, fill=gray!10, label=above:{}] (9)    at (12,-16)	{$ u_k^3 $ };
\node[state, fill=gray!10, label=above:{}] (10)    at (8,-20)	{$ \dots $};
\node[state, fill=gray!10, label=above:{}, accepting] (11)    at (4,-20)	{$ u_k^{p_k} $};

 \draw (0) edge  [] node [] {$a$} (1);
  \draw (1) edge [] node [] {$a$} (2);
 \draw (2) edge [] node [] {$a$} (3);
 \draw (3) edge  [dashed] node [] {$a$} (4);
 \draw (4) edge  [dashed] node [] {$a$} (5);
 \draw (5) edge  [] node [] {$a$} (1);
  \draw (0) edge  [] node [] {$a$} (7);
  \draw (7) edge [] node [] {$a$} (8);
 \draw (8) edge [] node [] {$a$} (9);
 \draw (9) edge  [dashed] node [] {$a$} (10);
 \draw (10) edge  [dashed] node [] {$a$} (11);
 \draw (11) edge  [] node [] {$a$} (7);
  \draw (0) edge  [dashed] node [] {$a$} (6);

\end{tikzpicture}
 \end{center}
 	\caption{An NFA recognizing $ \mathcal{L} = \{a^r \mid   (\exists  i \in \{1, \dots, k  \})(\text{$ p_i $ divides  $ r $})  \}$.}\label{fig:nfaexp}
\end{figure}

\paragraph*{\textbf{Proof}}
Consider the language:
\begin{equation*}
    \mathcal{L} = \{a^r \mid   (\exists  i \in \{1, \dots, k  \})(\text{$ p_i $ divides  $ r $})  \}.
\end{equation*}
The NFA in Figure \ref{fig:nfaexp} recognizes $ \mathcal{L } $ and it has $ 1+\sum_{i = 1}^k p_i $ states, so its width is at most $ \sum_{i = 1}^k p_i $ (because the initial can be compared with any other state).

\begin{figure}[h!] 
 \begin{center}
\begin{tikzpicture}[->,>=stealth', semithick, auto, scale=.3]
\tikzset{every state/.style={minimum size=0pt}}

\node[state,fill=gray!10, initial, accepting ] (0)    at (0,0)	{$ u_0 $ };

\node[state,fill=gray!10, label=above:{}] (1)    at (4,4)	{$ u_1 $ };
\node[state, fill=gray!10, label=above:{}] (2)    at (8,4)	{$ u_2 $};
\node[state, fill=gray!10, label=above:{}] (3)    at (12,0)	{$ u_3 $ };
\node[state, fill=gray!10, label=above:{}] (4)    at (8,-4)	{$ \dots $};
\node[state, fill=gray!10, label=above:{}] (5)    at (4,-4)	{$ u_\ell $};

 \draw (0) edge  [] node [] {$a$} (1);
  \draw (1) edge [] node [] {$a$} (2);
 \draw (2) edge [] node [] {$a$} (3);
 \draw (3) edge  [dashed] node [] {$a$} (4);
 \draw (4) edge  [dashed] node [] {$a$} (5);
 \draw (5) edge  [] node [] {$a$} (1);

\end{tikzpicture}
 \end{center}
 	\caption{The topology of all DFAs recognizing $ \mathcal{L} = \{a^r \mid   (\exists  i \in \{1, \dots, k  \})(\text{$ p_i $ divides  $ r $}.)\}$. Some states of the cycle are final.}\label{fig:dfaexp}
\end{figure}
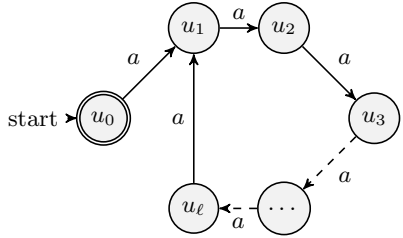

In order to prove that $ {\text width}^D(\mathcal L) \ge \prod_{i = 1}^k p_i $, consider any DFA $\mathcal A$ recognizing $\mathcal L$. Since $ \mathcal{A} $ is deterministic and $ \mathcal{L} $ is an infinite language, then $ \mathcal{A} $ must be like the one in Figure \ref{fig:dfaexp}, with a simple cycle of length - say - $ \ell $. Let us prove that $\prod_{i = 1}^k p_i$ divides $\ell$.
Consider an integer $m$ such that the word   $a^{s_0}$ reaches a state $ u $ inside the cycle, where $s_0=(\prod_{i=1}^kp_i)^m $. State $ u $ must be final because $ a^{s_0} \in \mathcal{L} $. If $s_1=s_0+\ell$, then also $a^{s_1} \in I_u $ and so $ a^{s_1} \in \mathcal{L} $. Without loss of generality, we can assume that $ p_1 $ divides $ s_1 $. Since $ p_1 $ divides both $ s_0 $ and $ s_1 $, we conclude that $ p_1 $ divides $ \ell $. If we now repeat this argument with $ (\prod_{i=2}^kp_i)^m $ and $ (\prod_{i=2}^kp_i)^m + \ell $, we obtain again that some $ p_i $ divides $ (\prod_{i=2}^kp_i)^m + \ell $. Since it cannot be $ p_i = p_1 $ (because $ p_i $ divides $ \ell $ but it does not divide $ (\prod_{i=2}^kp_i)^m $), we conclude without loss of generality that $ p_2 $ divides $ \ell $. Proceeding like this, we conclude that every $ p_i $ divides $ \ell $ and so $ \prod_{i = 1}^k p_i$ divides $\ell$. By Lemma \ref{lem:NFAcycle} we conclude that $\text{width}(\mathcal A)\geq \prod_{i = 1}^kp_i$, and so $ {\text width}^D(\mathcal L) \ge \prod_{i = 1}^k p_i $ being $ \mathcal{A} $ arbitrary.

\hfill $\Box$

\paragraph*{\textbf{Statement of Lemma \ref{lem:lower bound}.}}
There exist a regular language $ \mathcal{L} $, whose width can be chosen arbitrarily large, such that $width^{D}(\mathcal L) \geq e^{\sqrt{width^{N}(\mathcal L)}}$.

\paragraph*{\textbf{Proof}}
First, notice that if the deterministic width can be chosen arbitrarily large, then also the nondeterministic width can be chosen arbitrarily large by Corollary \ref{cor:widths}. Let $p_1, \dots, p_k$ be all primes no larger than a fixed $n$. The primorial function grows asymptotically as $\prod_{i = 1}^k p_i = e^{(1+o(1))n} \geq e^n$ and the sum of the primes no larger than $n$ grows asymptotically as  $\sum_{i = 1}^k p_i \in O(n^2/\log n)$ and it is, in fact, never larger than $n^2$.
Now, consider the language $\mathcal L$ of Lemma \ref{lem:det_non_width}.
Combining $width^{N}(\mathcal L) \leq n^2$ with $width^{D}(\mathcal L)  \geq e^n$, we obtain the claimed lower bound.
 \hfill $\Box$

\section{Proofs of Section \ref{sec:indexing}}

\paragraph*{\textbf{Statement of Lemma \ref{lem:nonemptyR}.}}
The output of Algorithm \ref{alg:R(A,u,v)} - that is, $ \mathcal R(\mathcal A,u,v) $ - is nonempty if and only if there exists a co-lex order  on $\mathcal A$ containing the pair $(u,v)$.  

\paragraph*{\textbf{Proof}}
If $\mathcal R(\mathcal A, u,v) \neq \emptyset$, we prove that     $\mathcal R(\mathcal A, u,v)$ satisfies properties  1 and 2 of Lemma  \ref{lem:transitive_closure}. Let $\rho$ be the incrementally-growing relation initialized at Line \ref{init rel} of Algorithm \ref{alg:R(A,u,v)}. Clearly, if $u,v\in Q$ are such that $\lambda(u)\prec \lambda(v)$ then the pair $(u,v)$ has been added to $\rho$ in line $7$ and thus belongs to the final relation $\mathcal R(\mathcal A, u,v)$. Hence,  the first  property      of Lemma  \ref{lem:transitive_closure}  is satisfied. As for the second, suppose $(u',v')\in \mathcal R(\mathcal A, u,v) $ and   $(u'',v'')$ is a pair such that $(u'',a,u)\in E_{\mathcal A},  (v'',a,v)\in E_{\mathcal A}$. Then, during some iteration of the \texttt{repeat} loop, at Line \ref{pop} the  pair $(u',v')$  was extracted from the stack, and,  since   $\mathcal R(\mathcal A, u,v) \neq \emptyset$, lines 11,12 where skipped and  the pair $(u'',v'')$ was added to $\rho$ on line 14.
  This proves that $\mathcal R(\mathcal A, u,v)$   satisfies  1 and 2 of Lemma  \ref{lem:transitive_closure}. Moreover,  the reflexive and transitive closure of $\mathcal R(\mathcal A, u,v)$ is antisymmetric: to see this it is sufficient to note that at Line \ref{acyclic} we return a nonempty relation if and only if $\rho$ is acyclic. 
  By Lemma \ref{lem:transitive_closure},
  it follows that the reflexive and transitive closure of $\mathcal R(\mathcal A,u,v)$ is a co-lex order containing the pair $(u,v)$.

  Conversely, if $\leq$ is a co-lex order containing the pair $(u,v)$, then Algorithm \ref{alg:R(A,u,v)} cannot return $\emptyset$ because, inductively, all pairs in $\rho$ are contained in $\leq$ as a consequence of co-lex axioms.
  
  \hfill $\Box$

\paragraph*{\textbf{Statement of Lemma \ref{lem:complexity}.}}
Algorithm \ref{alg:R(A,u,v)} can be implemented so that it terminates in $O(|Q|^4)$ time.

\paragraph*{\textbf{Proof}}
$\rho$ can be implemented as an adjacency matrix (a bitvector) with $|Q|^2$ entries so that membership and insert queries take constant time. $S$ is a standard stack with constant-time operations. 
There are at most $O(|Q|^2)$ possible pairs $(u,v)$ that can be inserted into $S$. Moreover,  no pair can be inserted more than once into $S$ since (1) upon insertion of a pair into $S$, we also insert it into $\rho$,  (2) a pair is not inserted into $S$ if it already belongs to $\rho$, and (3) we never remove pairs from $\rho$. At Line 10 we check all possible $a$-predecessors (at most $O(|Q|^2)$) of a pair extracted from $S$. To conclude, the acyclicity test at Line 17 can be solved in $O(|\rho|) = O(|Q|^2)$ time using Kahn's algorithm.
 \hfill $\Box$

\paragraph*{\textbf{Statement of Lemma \ref{lem:alg}.}}
Consider the relations $\rho_\exists$ and $\sim$.
\begin{enumerate}
    \item If $ (u, v) \in \rho_\exists $, then $ I_u \preceq I_v $;
    \item If $u \sim v$, then $ I_u = I_v $.
\end{enumerate}

\paragraph*{\textbf{Proof}}
\begin{enumerate}
\item If $ (u, v) \in \rho_\exists $, then  the   pair $(u,v)$ belongs to a co-lex order (Lemma \ref{lem:nonemptyR})  and $ I_u \preceq I_v $ follows from Corollary \ref{cor:setstrings}.
\item If $u\sim v$, then there exist $ z_1, \dots, z_k $ (with $ k \ge 0 $) such that $ (u, z_1) \in \rho_\exists $, $ (z_1, z_2) \in \rho_\exists $, $ \dots $, $ (z_k, v) \in \rho_\exists $, and there exist $ z'_1, \dots, z'_h $ (with $ h \ge 0 $) such that $ (v, z'_1) \in \rho_\exists $, $ (z'_1, z'_2) \in \rho_\exists $, $ \dots $, $ (z'_h, u) \in \rho_\exists $. By the previous point, we have $ I_u \preceq I_{z_1} \preceq \dots \preceq I_{z_k} \preceq I_v \preceq I_{z'_1}\preceq  \dots \preceq I_{z'_{h}} \preceq I_u  $, so we conclude $ I_u = I_v $ by Lemma \ref{lem:preceq}.
\end{enumerate}
 \hfill $\Box$

\paragraph*{\textbf{Statement of Lemma \ref{lem: triangle properties}.}}
The following properties hold:
\begin{enumerate}
    \item if $ u \trianglelefteq v $, then $ I_u \preceq I_v $;
    \item $ \trianglelefteq $ is a partial order;
    \item $ width (\trianglelefteq) \preceq width (\mathcal{A}) $;
    \item $ \trianglelefteq $ can be computed in $O(|Q|^6)$  time.
\end{enumerate}

\paragraph*{\textbf{Proof}}
\begin{enumerate}
    \item First notice that if $ (u, v) \in R $, then $ I_u \preceq I_v $ by Lemma \ref{lem:alg} and the definition of $ R $. Now, if $ u \trianglelefteq v $, then there exist $ z_1, \dots, z_k $ (with $ k \ge 0 ) $ such that $ (u, z_1) \in R $, $ (z_1, z_2) \in R $, $ \dots $, $ (z_k, v) \in R $, hence we conclude $ I_u \preceq I_{z_1} \preceq \dots \preceq I_{z_k} \preceq I_v $ and in particular $ I_u \preceq I_v $.
     
    \item We only have to prove that $ \trianglelefteq $ is antisymmetric. Assume, for contradiction, that $ q_i \trianglelefteq q_j $, $ q_j \trianglelefteq q_i $ and $ q_i \not = q_j $. This implies that there exist $ q_{i_1}, \dots, q_{i_k} $ (with $ k \ge 0 ) $ such that $ (q_i, q_{i_1}) \in R $, $ (q_{i_1}, q_{i_2}) \in R $, $ \dots $, $ (q_{i_k}, q_j) \in R $, and there exist $ q_{i'_1}, \dots, q_{i'_h} $ (with $ h \ge 0 ) $ such that $ (q_j, q_{i'_1}) \in R $, $ (q_{i'_1}, q_{i'_2}) \in R $, $ \dots $, $ (q_{i'_h}, q_i) \in R $. By Remark \ref{rem:path rho}, this implies that there exist two paths, one from $q_i$ to $q_j$  and one from $q_j$ to $q_i$, in $\rho_\exists$. 
    Then, $q_i, q_{i_1}, \dots, q_{i_k}, q_j$ belong to the same strongly connected component of $\rho_\exists$, thus they are also pairwise $\sim$-equivalent. Since $ (q_i, q_{i_1}) \in R $, $ (q_{i_1}, q_{i_2}) \in R $, $ \dots $, $ (q_{i_k}, q_j) \in R $, the definition of $R$ finally implies $i<i_1<i_2< \dots < i_k < j$. Similarly, we obtain $j < i'_1 < \dots < i'_h < i$: a contradiction.

    \item Let $ \le $ be a co-lex order on $ \mathcal{A} $ such that $ width (\le) = width (\mathcal{A}) $. If $ q_i $ and $ q_j $ are $ \le $-comparable, then $ q_i $ and $ q_j $ are $ \rho_\exists $-comparable (by the definition of $ \rho_\exists $ and Lemma \ref{lem:nonemptyR}). Hence $ q_i $ and $ q_j $ are $ R $-comparable: if $q_i\sim q_j$, then by definition of $R$, we have $(q_i,q_j) \in R$ if and only if $i<j$, while, if $q_i \not\sim q_j$ then $(q_i,q_j) \in R$ if and only if $(q_i,q_j) \in \rho_\exists$  and either $(q_i,q_j) \in \rho_\exists$ or $(q_j,q_i) \in \rho_\exists$ holds by assumption. In either case, $ q_i $ and $ q_j $ are $ R $-comparable.
    We conclude that $ u $ and $ v $ are $ \trianglelefteq $-comparable since $ \trianglelefteq $ is the reflexive and transitive closure of $R$. This implies $ width (\trianglelefteq) \le width (\mathcal{A}) $.
    \item The bottleneck is computing $\rho_{\exists}$: by Lemma \ref{lem:complexity}, we need to run Algorithm \ref{alg:R(A,u,v)} ($O(|Q|^4)$ time) for each pair of states. Once $\rho_\exists$ has been computed, computing the strongly connected components of $\rho_\exists$ only takes $|\rho_\exists| = O(|Q|^2)$ time. The final order $\trianglelefteq$ can easily be computed with a visit of the strongly connected components graph, in additional $O(|Q|^2)$ time.  
\end{enumerate}
 \hfill $\Box$

\paragraph*{\textbf{Statement of Lemma \ref{lem:equalrelation}.}}
Let $ \mathcal{A} $ be an NFA that admits the maximum co-lex order $ \le $. Then, $\trianglelefteq$ and $ \le $ are the same relation.

\paragraph*{\textbf{Proof}}
If  $(u,v)\in \rho_\exists$, then $u < v$, because there must exist a co-lex order $ \le' $ such that $ u <' v $, and $ \le $ is the maximum co-lex order. This implies that the equivalence classes of the relation $\sim$  from Definition \ref{def:rhoexists} are singletons. If it were not so, then there would exists $ u, v $ such that $ u \sim v $ and $ u \not = v $. From $ u \sim v $ we obtain that there exist $z_1, \ldots, z_k$ (with $ k \ge 0 $) such that $(u,z_1)\in \rho_\exists $, $ (z_1, z_2) \in \rho_\exists $, $ \ldots $, $ (z_k,v)\in \rho_\exists $, and there exist $z'_1, \ldots, z'_h$ (with $ h \ge 0 $) such that $(v,z'_1)\in \rho_\exists $, $ (z'_1, z'_2) \in \rho_\exists $, $ \ldots $, $ (z'_h,u)\in \rho_\exists $. As a consequence, we obtain $ u < z_1 < \dots < z_k < v $ and $ v < z'_1 < \dots < z'_h < u $, a contradiction.

First, assume that $ u < v $, and let us prove that $ u \trianglelefteq v $. From $ u < v $ it follows $ (u,v)\in \rho_\exists $. Since $u\neq v$ and $\sim$ classes are singletons, by Definition \ref{def:triangle} we get $(u,v)\in R$ so that $u\trianglelefteq v$ follows. Conversely, assume that $u \trianglelefteq v$ and $u\neq v$, and let us prove that $ u < v $. There must exist states $ z_1, \dots, z_k $ (with $ k \ge 0 $) such that $ (u, z_1) \in R $, $ (z_1, z_2) \in R $, $ \dots $, $ (z_k, v) \in R $. Since $\sim$-classes are singletons, then the definition of $R$ implies that $ (u, z_1) \in \rho_\exists $, $ (z_1, z_2) \in \rho_\exists $, $ \dots $, $ (z_k, v) \in \rho_\exists $, so $ u < z_1 < z_2 < \dots < z_k < v $ and in particular $ u < v $.
 \hfill $\Box$

\paragraph*{\textbf{Statement of Lemma \ref{lem:WPC}.}}
 Let $\leq$ be any partial order over the states $Q$ of an NFA $\mathcal A = (Q,s,\delta,F)$, such that $u \leq v \Rightarrow I_u \preceq I_v$. Then, $(Q,\leq)$ is weakly path coherent.
 
 \paragraph*{\textbf{Proof}}
 Suppose $u,v,w\in Q$, $u<v<w$ and $u,w\in \mathcal B(P)$.  We have to prove that $v\in \mathcal B(P)$. If $I_u=I_v$ or $I_w=I_v$ we are done.  Otherwise, by hypothesis 
 we have $I_u\prec I_v\prec I_w$. From $u,w\in \mathcal B(P)$ we know that there exists  $\alpha=\alpha'P\in I_u$ and  there exists  $\gamma=\gamma'P\in I_w$. If $\alpha \in I_v$ or $\gamma \in I_v$, then $v\in \mathcal B(P)$, and we are done. Otherwise, $\alpha\in I_u\setminus I_v$ and $\gamma \in I_w\setminus I_v$, so if we pick any $ \beta \in I_v $, from  $I_u\prec I_v \prec I_w$ it follows $\alpha \prec \beta \prec \gamma$ which implies $\beta =\beta' P$ and $v\in \mathcal B(P)$.
  \hfill $\Box$

 \paragraph*{\textbf{Statement of Lemma \ref{lem:DS}.}}

    There is a representation for arrays $OUT_i$ taking $O(|\mathcal A|)$ words of space and  simulating the two inner \texttt{for} loops of Algorithm \ref{alg:extension} in $O(\log(t\cdot \sigma))$ time.

\paragraph*{\textbf{Proof}}

The basic operation we need to speed up this step is the following: given a range $l,r$, find the triple $(a',j',q)$ in $OUT_i[l,r]$ such that (i) $a'=a$, (ii) $j'=j$, and (iii) $q$ is minimized. The solution for finding the maximum is symmetric, so we do not discuss it here. First, we concatenate all triples in each $OUT_i[k]$ in a single sequence $W_i$. A succinct bitvector with constant-time rank and select operations (for example, see \cite{raman2007succinct}) can be used to retrieve in constant time the interval $W_i[l',r']$ containing all the triples $OUT_i[l,r]$, given $l$ and $r$. Inside each list $W_i$, we represent each $(a',j',q)$ as a pair $(bin(a',j'),q)$, where $bin(a',j')$ is the integer obtained by concatenating the binary representations of $a'$ and $j'$, using a fixed number $h = |bin(a',j')| = O(\log(t\cdot \sigma))$ of bits. 
Finally, we build a wavelet tree \cite{grossi2003high,ferragina2009myriad} over each sequence $W_i$, treating the second component $q$ of each pair $(bin(a',j'),q)$ as satellite data: the wavelet tree has height $h$, and the leaf obtained by descending the tree by the binary sequence $bin(a',j')$ is a sequence of satellite data $W_{i,a',j'} = \langle q\ |\ (Q_i[k], Q_{j'}[q],a') \in E_{\mathcal A}  \rangle$. Most importantly, given indices $l,r$, by descending the wavelet tree by the sequence $bin(a',j')$ (in time $h \in  O(\log(t\cdot \sigma))$) starting from range $[l,r]$ on the wavelet tree's root,  we obtain the two indices $l', r'$ such that $W_{i,a',j'}[l',r'] = \langle q\ |\ (Q_i[k], Q_{j'}[q],a') \in E_{\mathcal A}\ \wedge \ l \leq k \leq r \rangle$. Finally, we build a range-minimum data structure \cite{fischer2010optimal} over the satellite data of each leaf $W_{i,a',j'}$ of the wavelet tree. Such a structure allows retrieving in constant time the minimum element in $W_{i,a',j'}[l',r']$, for any range $l',r'$.

 \paragraph*{\textbf{Statement of Theorem \ref{th:FM-index}.}}
 Let $ \mathcal{A} = (Q, s, \delta, F) $ be an NFA with $p = width(\mathcal A)$. In polynomial time we can build a data structure taking $O(|\mathcal{A}|)$ words of space that, given a query string $\alpha\in \Sigma^\patlen$,  supports the following operations in $O(\patlen\cdot p^2 \cdot \log(p\cdot \sigma))$ time:  
\begin{itemize}
    \item[(i)] Count the number of states reached by a path labeled $\alpha$.
    \item[(ii)] Return unique identifiers for the states reached by a path labeled $\alpha$. 
    \item[(iii)] Decide whether $\alpha \in \mathcal L(\mathcal A)$.
\end{itemize}
 
\paragraph*{\textbf{Proof}}
 As proved in Lemma \ref{lem:DS}, a single character-extension step takes $O(p^2\cdot \log(p\cdot \sigma))$ time with our data structures. 
 Given the representation $(l_i,r_i)_{i=1}^t$ of $\mathcal B(P)$, 
 query (i) (counting) amounts to returning $\sum_{i=1}^t (r_i-l_i+1)$. Query (ii) amounts to returning any node identifier, stored as satellite data $X_i[l_i, r_i]$, associated with each  linear component $Q_i$. Finally, query (iii) can be solved by simply marking in a bitvector every final state (in the order of the linear components). By augmenting each bitvector with rank and select functionality  \cite{raman2007succinct}, one can discover in constant time per linear component whether any state in $\mathcal B(P)$ is final.
  \hfill $\Box$

\bibliographystyle{plainurl}
\bibliography{mainbiblio}

\end{document}